\documentstyle[aps,prl,epsf,multicol]{revtex}
\begin{document}

\title{Active Width at a Slanted Active Boundary in Directed Percolation}
\author{Chun-Chung Chen${^1}$, Hyunggyu Park$^{1,2}$, Marcel den Nijs${^1}$}
\address{$^1$ Department of Physics, University of Washington,
         Seattle, Washington 98195-1560, U.S.A.}
\address{$^2$ Department of Physics, Inha University,
         Inchon 402-751, Korea}
\maketitle

\begin{abstract}
The width $W$  of the active region around an active moving wall
in a directed percolation process diverges at the percolation threshold $p_c$ 
as  $W\simeq A \epsilon^{-\nu_\parallel} \ln(\frac{\epsilon_0}{\epsilon})$,  
with $\epsilon= p_c-p$, $\epsilon_0$ a constant, and $\nu_\parallel=1.734$
the critical exponent of the characteristic time 
needed to reach the stationary state $\xi_\parallel\sim
\epsilon^{-\nu_\parallel}$.
The logarithmic factor arises from screening of statistically 
independent needle shaped sub clusters in the active region.
Numerical data confirm this scaling behaviour.
\end{abstract}

\begin{multicols}{2}
\section{Introduction}

Directed percolation (DP) has emerged as one of the generic absorbing 
state type dynamic processes. 
It describes epidemic processes, e.g., forest fires and
various types of surface catalysis processes
\cite{KinzelBook,Schlogl,Grassberger79,Grassberger78,Ziff}.
Such processes include a so-called absorbing state,
typically the vacuum, from which it can not escape.
The relevant tunable parameter is the propagation probability $p$. 
The system undergoes a phase transition 
from the absorbing phase at small $p$, where the stationary state is the 
absorbing state, into an active stationary phase at large $p$, 
where the system refuses to die.
The scaling properties at DP  dynamic phase transitions are 
known for almost two decades, and it's now realized
that DP critical behaviour is the generic
universality class for dynamic absorbing state type processes
\cite{KinzelBook}.

At DP type critical points the equilibration time $\xi_\parallel$ diverges.
It scales as $\xi_\parallel\sim \xi_\perp^z$ compared to the 
spatial correlation length $\xi_\perp$, with
dynamic exponent $z = 1.581$\cite{Jensen}.
For example, starting from a single seed, the
survival probability  obeys the scaling form
\begin{equation}
P_s(\epsilon, t) = 
b^{-x_s} P_s( b^{1/\nu_\perp} \epsilon, b^{-z} t)
\label{surv1}
\end{equation}
with $\epsilon=p_c-p$ the distance from the critical point.
This leads to 
\begin{equation}
P_s\sim \epsilon^\beta \exp(\frac{-t}{\xi_\parallel}),
\label{surv2}
\end{equation}
with exponent $\beta=x_s \nu_\perp$.
The exponential factor reflects that deep inside the absorbing phase
$P_s$ decays exponentially in time.
The equilibration time diverges at the DP critical point 
as $\xi_\parallel\sim \epsilon^{-\nu_\parallel}$ with
$z=\nu_\parallel/\nu_\perp$.
At $p_c$  the survival probability decays as a powerlaw,
$P_s(t)\sim t^{-\delta}$ with $\delta = x_s/z =\beta/\nu_\parallel$.

A recent direction  of research in this topic
concerns the scaling properties near boundaries
\cite{Essam,Frojdh,Lauritsen98,Lauritsen}.
Those studies address absorbing and reflective walls.
The scaling properties are modified by surface type critical exponents.
In particular, the survival probability for a seed near the boundary obeys 
the same scaling form as above, but with a new interface critical exponent $x$,
and therefore a modified value for $\beta$.

In this study we discuss the scaling properties near active boundaries.
Consider a stationary active vertical wall in the system.
All sites in the wall are alive.
The critical exponent $\beta$ is not an issue, because the 
system remains active near the wall for all $p$.
However, in the absorbing phase
the cloud of active sites near the wall has a specific  stationary state
width, which is expected to diverge as 
$W \sim \xi_\perp\sim \epsilon^{-\nu_\perp}$.
Widths like this diverge with  bulk exponents.

\begin{figure}
\centerline{\epsfxsize=8cm \epsfbox{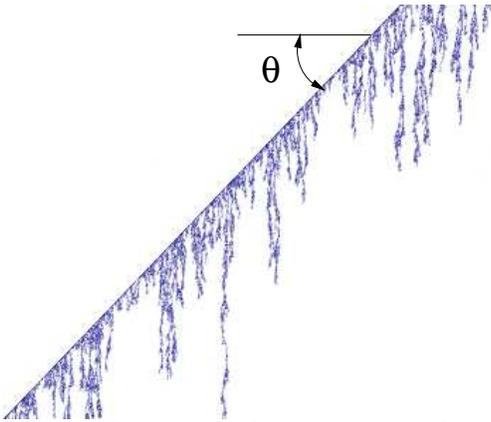}}
\caption{The curtain of active sites at the active slanted boundary}
\label{curtain}
\end{figure}

Assume that this  wall is slanted, with an arbitrary angle 
$\theta \neq 90^\circ$ with respect to the horizontal direction
(see Fig.\ref{curtain}).
In the space-time interpretation of the configurations, 
the wall moves with a constant velocity.
It acts as a slanted active curtain rod.
A curtain of active sites hangs down from it 
as illustrated in Fig.\ref{curtain}.
For $p<p_c$ the curtain has a finite width $l_\perp$ and 
length $l_\parallel = l_\perp \tan(\theta)$.
 
In this study we address how the stationary state
width of this slanted curtain scales near the DP critical point.
Naively this seems a simple question.
One would expect that the curtain width diverges 
with the same exponent as the equilibration time scale, 
$W\sim \epsilon^{-\nu_\parallel}$,
i.e., with the same exponent as the length of a curtain hanging down from an
horizontal curtain rod  ($\theta=0$).
The latter is equivalent to asking for the survival probability 
in the set-up without any walls 
where all sites are active in the initial state. 

This expectation is based on the anisotropic scaling properties. 
Consider a system with a rod at angle $\theta\neq 0$.
The horizontal  and vertical bulk lengths diverge with different exponents, as
$\xi_\parallel\sim \xi_\perp^z$.
Therefore, a system at $p_c-p=\epsilon$ and wall angle $\theta$
is equivalent by renormalization  to a system with  a smaller  wall angle
$\theta^\prime$ at  $\epsilon^\prime= b^{-1/\nu_\perp}\epsilon$ with 
$\tan(\theta^\prime)\simeq b^{z-1} \tan(\theta)$.
The scaling properties of $W$ should not depend on the angle $\theta$,
since the rod renormalizes towards the horizontal position.
We should expect the same scaling behaviour as at $\theta=0$.
However, a recent numerical study \cite{Park} seems to contradict this.

Kwon {\it et. al.}\cite{Park} studied a model with two absorbing states.
It undergoes a dynamic phase transition which belongs to Directed Ising (DI) 
universality class when the two absorbing states are symmetric,
and belongs to the Directed Percolation (DP) universality class when
a symmetry breaking field is introduced.
They studied the interface dynamics of the active domain between two
asymmetric absorbing states.
As one absorbing state dominates over the other,
the interface is driven into the unpreferred absorbing region with a
constant velocity.
Therefore they expected the width of the active domain
to scale  like the horizontal width of the active curtain
in the above setup for ordinary DP models.
A simple power-law fit of their data suggests that the active domain width
scales as
$W\sim \epsilon^{-x}$ with $x\simeq 2.00(5)$, which does not
agree with the DP exponent $\nu_\parallel\simeq 1.734$. 

In this paper we address the same issue more directly.
We insert  a slanted active wall into the most basic model for DP, 
the one studied originally by Kinzel\cite{Kinzel,Domany}, see section 2.
We find a similar anomalous value for the width exponent.
$W\sim \epsilon^{-x}$  scales as $x\simeq 1.95(5)$.
In section 3 we  develop a qualitative scaling theory.
It predicts that the curtain width scales with the conventional 
exponent  $\nu_\parallel$ but with an additional 
logarithmic factor as $W\simeq A \epsilon^{-\nu_\parallel}
\ln(\frac{\epsilon_0}{\epsilon})$.
In section 4 we show that the numerical Monte Carlo data fits this form well.
In section 5 we illustrates how DP type processes with slanted
walls can be studied  in the Master equation formalism.
Our finite size scaling (FSS) results, using  
exact numerical enumeration of the eigenvalue spectrum,
show that at $p_c$ the width of the slanted curtain diverges as $W\sim L^z$
with system size. This confirms the absence of a new independent exponent.
The logarithmic factor arises only in the $\epsilon$ dependence.
 
\section{ Numerical Results for the Curtain Width.}

\begin{figure}
\centerline{\epsfxsize=8cm \epsfbox{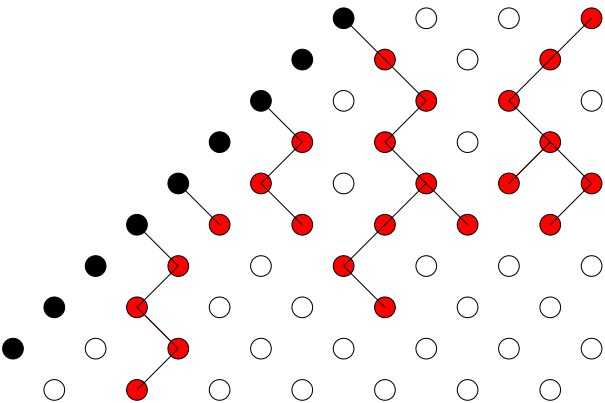}}
\caption{Lattice structure near the active boundary}
\label{lattice}
\end{figure}

Consider the square space-time lattice shown in Figure \ref{lattice}.
All bonds run under 45 degrees.
The black (open) circles represent the active (inactive) sites.
Time evolves from top to bottom in half units $t \to t+\frac{1}{2}$.
Bonds between nearest neighbor sites at $t$ and $t \to t+\frac{1}{2}$
are being created  with probability $p$ but only if the upper site is active.
Each bond activates the lower site. 
Kinzel studied this model in detail with Master equation type FSS
in the early eighties \cite{Kinzel}. 
The critical exponents and the location of the DP transition are known
quite accurately.
For example, the latest series expansion results put the DP phase
transition at $p_c\approx 0.6447$\cite{Jensen}.

We modify the  boundary conditions in this model to accommodate an active wall.
The lattice is semi-infinite, bound to the left  by the wall,
which runs away under $\theta=45^\circ$ as shown in Fig.~\ref{lattice}.
$45^\circ$ is its natural angle for the curtain rod for this specific lattice.
We can restrict ourselves to this angle because the scaling properties of the 
curtain width should not depend on the angle according to the 
anisotropic scaling argument outlined above.
Moreover, the angle is a continuous parameter 
in the model by Kwon {\em et. al.}\cite{Park} and their 
results show no angle dependence.

\begin{figure}
\centerline{\epsfxsize=8cm \epsfbox{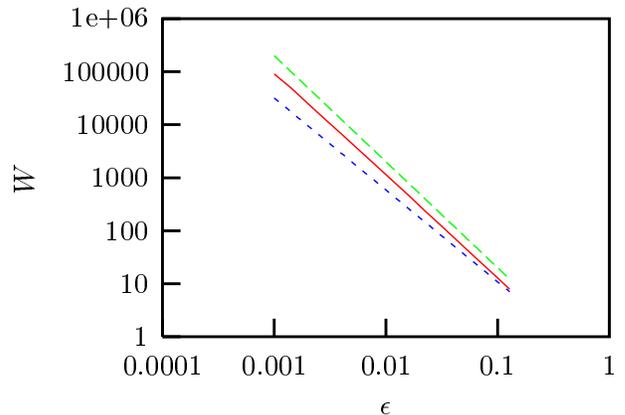}}
\caption{Log plot of active width versus $p_c-p$ from straight Monte Carlo
simulations on unlimited system sizes. The solid line represents the data.
The dashed straight lines of slopes $-2$ and $-1.734$ are guides to the
eyes.} 
\label{width}
\end{figure}

\begin{figure}
\centerline{\epsfxsize=8cm \epsfbox{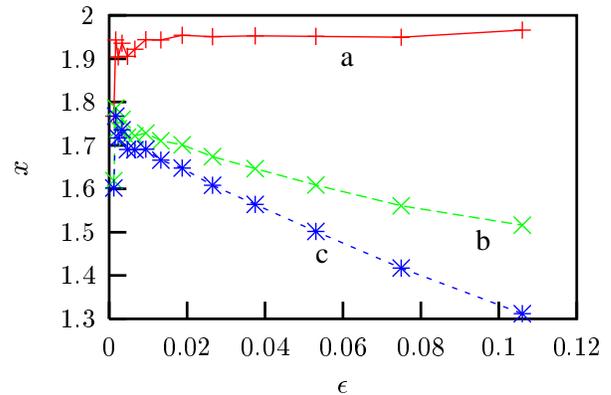}}
\caption{ Estimates for the  active width exponent, $x$. In fit (a), $W$ is
assumed to scale as $W \sim \epsilon^{-x}$, in (b), as $W\sim\epsilon^{-x}
\ln\epsilon$ and in (c), as $W\sim\epsilon^{-x}(\ln\epsilon+\ln 2)$.}
\label{width-exp}
\end{figure}

We perform Monte Carlo simulations
with as initial configuration an active wall in an inactive bulk.
The horizontal curtain width is defined as the distance of the last
active site from the rod in each time slice.
For $p<p_c$, the width grows initially approximately linear in time,
until it saturates at the stationary state value which varies with
$\epsilon=p_c-p$. 
Figure \ref{width} shows the active
width versus $\epsilon$ on a logarithmic scale.
The line is quite linear over the two decades shown.
The slope is clearly distinct from the expected value
$\nu_\parallel\approx 1.734$ and close to the value
found by Kwon {\em et. al.}\cite{Park}.
In Fig.\ref{width-exp} we perform a more careful FSS analysis to the same
data. We fit the numerical data from two nearby points,
$\epsilon_2=\sqrt2 \epsilon_1$,
to the form $W\simeq a\epsilon^{-x}$ and plot $x$ as
a function of $\epsilon$, the exponent
$x$ appears to be around $1.95$. 
This fit is remarkably stable, and shows virtually no power law type
corrections to scaling.
Taken out of context it is strongly suggestive of a new
independent critical exponent.
The other curves in Fig.\ref{width-exp} relate to the FSS
analysis assuming an additional
logarithmic factor as discussed in the next two sections.
 
\section{Independent Cluster Approximation}

Figure \ref{curtain} shows a typical curtain configuration 
in a Monte Carlo simulation
at a $p$ just below the percolation threshold $p_c$.
The most striking features are the needles in the curtain.
Isolated clusters are expected to be needle like. 
The correlation length in the time direction diverges faster 
than in the spatial direction, as $\xi_\parallel\sim \xi_\perp^z$.
Therefore, active clusters (when grown from a single seed) 
become needle shaped near the percolation threshold. 
Figure \ref{curtain} gives the impression that close to $p_c$, 
the curtain consists of a set of weakly interacting needle shaped clusters
when viewed from length scales larger than $\xi_\perp$.

In this section we pursue the implications of the assumption that such
needles are completely uncorrelated.
In that approximation the probability that the curtain
extends over a horizontal distance
$l$ is given by the probability that a  needle longer than
$\tau = l \tan (\theta)$  hangs down from the curtain rod vertically
above that site.
Let $P$ be that probability.
It must have the same form as the survival probability 
from a single seed, Eq.(\ref{surv2}),
The actual value of the exponent $\beta$ turns out to be irrelevant 
in this section, but it must be identical to the single seed value,
according to a time reversal symmetry argument \cite{CCC}.

The spatial coordinate needs to be coarse grained, because the
needles can only be uncorrelated beyond the horizontal correlation length
$\xi_\perp\sim \epsilon^{-\nu_\perp}$.
Define $n= x/ \xi_\perp$ as the coarse grained discrete spatial coordinate
and recall that $t= x \tan(\theta)$ is the corresponding vertical distance
from the curtain rod to the same point.
The probability for the curtain to have width $n$ factorizes in the
independent needle approximation as
\begin{equation}
P_w(n)=P (n) \prod_{n^\prime>n} \bigg[ 1- P ( n^\prime ) \bigg].
\label{Pw}
\end{equation}
This equation can be rewritten into a derivative form 
\begin{equation}
\frac{P_w(n+1)-P_w(n)}{P_w(n+1)} = \frac{P (n+1)-P(n)[1-P(n+1)]}{P (n+1)}
\label{Pwd}
\end{equation}
The maximum of the distribution obeys the relation
\begin{equation}
P_w(\tilde n-1)=P_w(\tilde n)
\end{equation}
and can be written as 
\begin{equation}
\frac{1}{P(\tilde n)}-\frac{1}{P(\tilde n-1)}=1.
\label{Pmax}
\end{equation}
Assume that $P$ has the same asymptotic form as the single seed
survival probability, in Eq.(\ref{surv2}),
and that the maximum of the distribution occurs in this range of $n$.
The transformation to the coarse-grained
$n= x/ \xi_\perp\sim x \epsilon^{\nu_\perp}$
variable changes the critical exponent inside the exponential factor
\begin{equation}
P\simeq B \epsilon^\beta e^{- b n \epsilon^\Delta}
\label{Psurv}
\end{equation}
with $\Delta = \nu_\parallel-\nu_\perp$, and $b \sim \tan(\theta)$.
Inserting this form into Eq.(\ref{Pmax}) leads to 
\begin{equation}
1- e^{-b  \epsilon^\Delta} = B \epsilon^\beta  e^{-b \tilde n  \epsilon^\Delta}
\end{equation}
and, after expanding  the exponential on the left hand side, to 
\begin{equation}
b \tilde n  |\epsilon|^\Delta \simeq  \ln(\frac{B}{b}) +
 (\beta-\Delta) \ln(\epsilon) 
\end{equation}
In original units this reads
\begin{equation}
\tilde W  \simeq A  \epsilon^{-\nu_\parallel}\ln(\frac{\epsilon_0}{\epsilon})
\label{W-scaling}
\end{equation}
The characteristic probability depends on the wall angle as 
$\epsilon_0\sim 1/\tan(\theta)$.
The most probable width $\tilde W$ scales with the expected exponent
$\nu_\parallel$ but contains an additional logarithmic factor.

Asymptotically the most probable and the average widths coincide.
Eq.(\ref{Pwd}) can be approximated in the continuum limit as
\begin{equation}
\frac{1}{P_w} \frac{d P_w}{d n} = 1-\frac{P(n)}{P(n+1)} + P(n)
\end{equation}
Close to $p_c$ and for large $n$, where $P$ obeys Eq.(\ref{Psurv}),
we can integrate this
\begin{eqnarray}
P_w (n) &
\sim & \exp[(1-e^{b\epsilon^\Delta}) n -\frac{B}{b}\epsilon^{\beta-\Delta}
 e^{-bn \epsilon^\Delta}] \\ \nonumber
 & \sim & e^{b n \epsilon^\Delta}\exp[ -\frac{B}{b}\epsilon^{\beta-\Delta}
 e^{-bn \epsilon^\Delta}]
\end{eqnarray}
This distribution decays exponentially on both sides of the most probable value
and becomes sharp at the critical point, $\epsilon\to 0$. 
We checked explicitly that the most probable and
average coincide in this limit, 
and scale asymptotically with the same logarithmic factor, as in
Eq.(\ref{W-scaling}).

\section{Logarithmic Corrections to Scaling Analysis. }

The logarithmic factor in the
independent  needle approximation formula for the curtain width 
\begin{equation}
W(\epsilon)  \simeq  A  \epsilon^{-\nu_\parallel}
 \ln(\frac{\epsilon_0}{\epsilon})
\label{W-log}
\end{equation}
does not change the asymptotic exponent. It is still equal to 
$\nu_{\parallel}$. However  the finite size scaling (FSS) approach
to this value is very singular.
A conventional FSS analysis involves the construction of approximants for 
the critical exponent $x$ by fitting the values of $W$ at to
nearby $\epsilon$ to 
a pure power law form, $W\sim \epsilon^{-x}$. This is equivalent to defining
$x(\epsilon)$ as a derivative and yields for the above logarithmic form 
\begin{equation}
x = -\frac{\epsilon}{W} \frac{d W}{d\epsilon} 
  =\nu_\parallel + \frac{1}{\ln(\frac {\epsilon_0}{\epsilon})}
\end{equation} 
This function approaches $\nu_\parallel$ in a singular manner.
In the interval $0.01<\epsilon/\epsilon_0<0.3$,
$x$ seems to converge convincingly with a linear correction to scaling term  
to an effective exponent  which is about $0.2$ too large.
One would have to go to extremely small $\epsilon$'s to see
the true convergence. 
The power law fit in Fig.\ref{width-exp} shows signs of this.

The two other curves in Fig.\ref{width-exp} 
show the FSS estimates for the exponent $\nu_\parallel$
according to the form Eq.(\ref{W-log}) with $\epsilon_0=1$ or $\epsilon_0=0.5$.
$\epsilon_0$ is unknown, but likely of order one.
Both curves converge towards the conventional value $\nu_\parallel=1.734$.
This is strong evidence for the presence of the logarithmic factor.

\section{finite size scaling at the percolation threshold}

The logarithmic factor originates from the screening  of independent needles. 
It should not play a role in the FSS at the percolation threshold itself, 
because there $\xi_\perp$ diverges,
and the independent needle concept becomes meaningless.

So the curtain width must scale as $W\sim L^z$ at $p_c$,
if it is really true that no independent new exponent is involved.
To confirm this we present in this section numerical data from 
master equation type finite size scaling  using exact enumeration.
We also performed Monte Carlo simulations but prefer to present our
master equation data since this method requires a technical novelty.

A moving wall is inconvenient in simulations.
The lattice is finite by necessity and the moving wall requires 
a much bigger lattice than the one actually used by the process. 
This is a handicap in particular for master equation calculations
where one evaluates the rate at which the stationary state is being reached
by letting time go to infinity at each lattice size $L$.
Those systems sizes are typically small, $L\leq 20$ in our case, because
phase space scales exponentially with $L$.
Compared to MC simulations the master equation method trades system size for
numerical accuracy, and the ability to perform a detailed corrections
to scaling analysis. The accuracy of the two methods is typically
comparable, except for specific issues, like the logarithmic
factor in the previous sections, which
require intrinsic large lattice sizes.

The solution to the moving wall problem
is to distinguish between the time and spatial directions of the 
dynamic process, $\hat e_\perp$ and $\hat e_\parallel$,
and the ones used in the master
equation.  There is no need for them to coincide.
We choose a set-up where the master equation's time and space  directions
are redirected in the following manner.
Lines of constant time are parallel to $\hat e_\parallel- \hat e_\perp$,
such that the  moving wall coincides with the $t=0$ line.
Lines of constant position are parallel to the x-axis, which 
in the dynamic process represented lines of constant time.

The following skewed dynamic rule implements this pace-time rotation.  
Consider a square space time lattice
(Fig.\ref{lattice} rotated over 45 degrees).
Each site in the master equation time slice $t$ 
is updated sequentially from right to left.
The probability for site $x$ at time $\tau$ to be active depends on whether 
site $x-1$ was active at the previous time $t-1$
and/or at this moment in time, $t$.
This set-up  requires screwed boundary conditions.
The forest fire runs under an angle.
In this new interpretation the active wall represents a fully active 
initial configuration.

The energy gap in the spectrum of the time
evolution operator (transfer matrix),
is related to the curtain width in the following manner.
Let $|I\rangle$ be the initial state of the master equation,
$|0\rangle$ the absorbing state,
and $\hat T$ be the transfer matrix.
The stochastic nature of the transfer matrix implies that the disordered state
$|D\rangle$ is a left eigenvector with eigenvalue $\lambda_0=1$.
Define $\hat a_x$ as the projection operator which returns one (zero)
when site $x$ active (inactive).
The curtain width is associated with the 
probability distribution for site $x$ to be active at time $t$ but
after that never again. This takes the form
\begin{equation}
P(x,t) = 
\lim_{t_F\to \infty}
\langle D |~ [(1-\hat a_x)\hat T]^{t_F-t}~ \hat a_x \hat T^t | I \rangle 
\end{equation} 
The operator $(1-\hat a_x)\hat T$ has $\lambda_0=1$ as largest eigenvalue since
\begin{equation}
(1-\hat a_x)\hat T] |0\rangle = |0\rangle 
\end{equation} 
and because attaching a projection operator to $\hat T$ can not result in an
eigenvalue larger than the largest in  $\hat T$.
Let $\langle L_x|$ be the corresponding left eigenvector
(which can be evaluated  numerically).
Inserting this leads to 
\begin{eqnarray}
P(x,t) & = & \langle L_x| \hat a_x \hat T^t | I \rangle \\ \nonumber
 & \simeq & \langle L_x| \hat a_x|\lambda_1 \rangle \lambda_1^t
 \langle \lambda_1 | I \rangle  \\ \nonumber
 & \sim & \exp[-t/\xi_t]
\end{eqnarray} 
with $\xi_t= \log(\lambda_1)$ and $\lambda_1$ the next
largest eigenvalue of $\hat T$.

\begin{figure}
\centerline{\epsfxsize=8cm \epsfbox{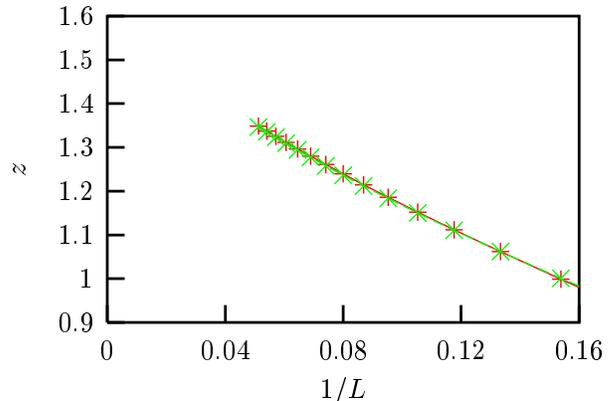}}
\caption{ Finite size scaling exponent $z$ for ($\times$) the characteristic
active width, $W\sim L^z$, and for ($+$) the time to reach the stationary
state, $t\sim L^z$, at the percolation threshold in the transfer
matrix set-up. The data virtually coincide}
\label{trans}
\end{figure}

This illustrates  that the curtain width scales in the same manner as the 
the characteristic time $\xi_t$  needed to reach the stationary
state, when the latter is measured in this space-time
twisted coordinate system.
Fig.\ref{trans} shows the FSS estimates for the dynamic exponent $z$
according to $\xi_t\sim L^z$ and $W\sim L^z$. 
Both converge clearly to the DP dynamic exponent $z=1.58$. 
This confirms that no new independent curtain width exponent is presents.
 
\section{Final Remarks}

The analysis presented in this paper explains the anomalous scaling of 
the width of the slanted curtain boundary in DP type processes.
The needles screen each other, and that leads an extra
logarithmic factor according to the independent needle approximation.
Our numerical data confirm the validity of this assumption.

The same mechanism must apply to other dynamic processes,
like directed Ising type absorbing state dynamics,
and also to other quantities.
Consider the following example.
Directed percolation describes epidemic growth processes without immunization,
where the probability to be sick at time $t+1$ requires that you  yourself
or at least  one of your neighbours is already sick at time $t$. 
Consider an initial condition that everybody is sick at time $t=0$.
A stationary local observer will conclude that below the percolation threshold
the life time of the epidemic scales as $t\sim \epsilon^{-\nu_\parallel}$.
A moving observer concludes it diverges faster, as 
$t\sim \epsilon^{-\nu_\parallel} \log(\frac{\epsilon_0 }{\epsilon})$.

This research is supported by NSF grant DMR-9700430,
by the Korea Research Foundation (98-015-D00090),
and by an Inha University research grant (1998).

\narrowtext

\end{multicols}

\begin{thebibliography}{99}
\bibitem{KinzelBook} W. Kinzel 1983 {\it Percolation Structures and Processes}
 (Ann. Isr. Phys. Soc. 5) ed G. Deutscher, R. Zallen
 \& J. Adler (Bristol:Hilger)
\bibitem{Schlogl} F. Schlogl, Z. Phys. 253 (1972) 147
\bibitem{Grassberger79} P. Grassberger \& A. de la Torre, Ann.
 Phys. NY 122 (1979) 373
\bibitem{Grassberger78} P. Grassberger \& K. Sundermeyer, Phys. Lett.
 77B (1978) 220
\bibitem{Ziff} R. M. Ziff, E. Gulari \& Y. Barshad, Phys. Rev. Lett.
 56 (1986) 2553
\bibitem{Jensen} I. Jensen, J. Phys. A29 (1996) 7013
\bibitem{Essam} J. W. Essam, A. J. Guttmann, I. Jensen \&
 D. TanlaKishani, J. Phys. A29 (1996) 1619
\bibitem{Frojdh} P. Frojdh, M. Howard \& K. B. Lauritsen, J. Phys. A31
 (1998) 2311
\bibitem{Lauritsen98} K. B. Lauritsen, P. Frojdh \& M. Howard, Phys. Rev.
 Lett. 81 (1998) 2104
\bibitem{Lauritsen} K. B. Lauritsen, K. Sneppen, M. Markosova \& M. H.
 Jensen, Physica A247 (1997) 1
\bibitem{Park} S. Kwon, W. Hwang \& H. Park, Phys. Rev. E 59 (1999) xxxx.
\bibitem{Kinzel} W. Klein \& W. Kinzel, J. Phys. A14 (1981) L163
\bibitem{Domany} E. Domany \& W. Kinzel, Phys. Rev. Lett. 53 (1984) 311
\bibitem{CCC} Chun-Chung Chen, unpublished, based on:
 J. L. Cardy \& R. L. Sugar, J. Phys. A13 (1980) L423, and
 H. Hinrichsen \& H. M. Koduvely, Eur. Phys. J. B5 (1998) 257.
\end{thebibliography}
\end{document}